\documentclass[aps,twocolumn, prl]{revtex4-2}

\usepackage{graphicx}  
\usepackage{dcolumn}   
\usepackage{bm}        
\usepackage{amsmath, amssymb, hyperref}   
\usepackage{physics}
\usepackage[dvipsnames]{xcolor}

\def\tr{\mrm{tr}}

\def\gge{\text{$Q$ grand-canonical ensemble }}

\renewcommand\({\begin{equation}}
\renewcommand\){\end{equation}}

\newcommand{\al}[1]{\begin{aligned}#1\end{aligned}}

\newcommand{\mrm}[1]{\mathrm{#1}}

\begin{document}

\title{Breaking a conservation law enables steady-state entanglement out of equilibrium}
\author{Vince Hou}
\author{Eric Kleinherbers}
\author{Shane P. Kelly}
\author{Yaroslav Tserkovnyak}
\affiliation{Department of Physics and Astronomy and Bhaumik Institute for Theoretical Physics, University of California, Los Angeles, California 90095, USA}
\date{\today}

\begin{abstract}
We show how entangled steady states can be prepared by purely dissipative dynamics in a system coupled to a thermal environment. 
While entanglement is hindered by thermalization when the system and environment exchange a conserved quantity, we demonstrate that breaking this conservation law through the system-environment interaction drives the system to a nonequilibrium steady state. Such an interaction will generate multiple competing equilibration channels, effectively mimicking baths at distinct chemical potentials. 
When the environment also supports long-range correlations, these channels mediate nonlocal dissipation capable of generating entanglement. 
We illustrate the scheme in a model of two nitrogen-vacancy (NV) centers weakly coupled to a spin-pumped  magnet, where tuneable magnon excitations enable steady-state entanglement over finite distances.
Our results identifies a general mechanism for dissipative entanglement generation, rooted in the conservation structure and environmental correlations rather than fine-tuned coherent control or active driving.

\end{abstract}

\maketitle

\begin{figure}[t!]
    \centering
    \includegraphics[width=0.49\textwidth]{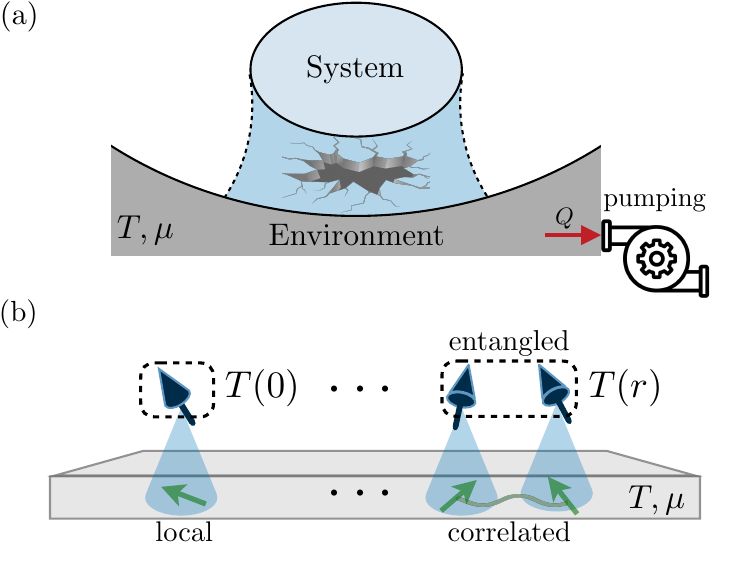}
    \caption{(a) The environment is prepared in a $Q$ grand-canonical ensemble at $Q$-chemical potential $\mu$ and temperature $T$.
    Breaking the conservation law of $Q$ via the coupling between system and environment (indicated by the crack) enables a genuine nonequilibrium steady state. For $\mu<0$, the quantity $Q$ is continuously injected into the environment, which attempts to restore equilibrium at $\mu=0$. Maintaining the environment at finite $\mu<0$ is achieved  via pumping.    
    (b) Implementation with NVs coupled via dipole-dipole interaction to a magnetic environment with temperature $T$ and spin accumulation $\mu$.
    To achieve steady-state entanglement, we need both local and correlated dissipation processes that are described by distinct effective temperatures, $T(0)$ and $T(r)$, see Eq.~\eqref{eq:T}.
    }
    \label{fig:1}
\end{figure}

\textit{Introduction}|Engineering entangled steady states is a stepping stone towards quantum applications such as quantum sensing \cite{RevModPhys.89.035002,PhysRevLett.118.040801}, quantum computation \cite{Gottesman1999,PhysRevLett.117.020402}, and quantum communication \cite{RevModPhys.86.419,PhysRevLett.126.250501}. Many entanglement preparation schemes seek to minimize unwanted noise by constructing well-isolated quantum systems \cite{Srinivas2021,Dolde2014}. Instead of secluding the system from its surroundings, other strategies utilize engineered dissipation to generate entangled states \cite{Verstraete2009}. In the weak-coupling regime, independent qubits exhibit transient entanglement when coupled to a generic thermal reservoir~\cite{PhysRevB.106.L180406}. However, the system state then necessarily evolves into a Gibbs ensemble when the environment is in thermal equilibrium \cite{Breuer, Schaller2014}. Deviation from a Gibbs steady state often requires nontrivial system designs, e.g., active driving \cite{PhysRevX.6.031036}, sophisticated interactions between the system's constituents \cite{PhysRevLett.128.080502,PhysRevB.105.L140301}, or far-from-equilibrium environments, such as waveguides \cite{PRXQuantum.5.030346}, and multi-reservoir engineering \cite{PhysRevLett.77.4728, PhysRevLett.134.176703,Khandelwal2025}. In this Letter, we demonstrate another perspective on dissipative dynamics: 
when a quantum system is weakly coupled to a thermal environment with temperature $T$ and chemical potential $\mu$ for quantity $Q$, we can prevent thermalization by breaking the conservation law of $Q$, see Fig.\ref{fig:1}a.

Inspired by recent developments in spin-optics interfaces~\cite{PhysRevX.2.011006,Chatterjee2021}, we illustrate our scheme by considering NV centers weakly coupled to a spin-pumped magnetic material~\cite{10.1093/oso/9780198787075.003.0008,RALPH20081190}, as shown in Fig.\ref{fig:1}b. 
The magnetic environment conserves the spin ($Q=S_z$), and is described by a spin grand-canonical ensemble with temperature $T$ and spin chemical potential $\mu$, better known as the spin accumulation~\cite{10.1093/oso/9780198787075.003.0008}.
If the coupling to the NVs conserves the total spin, standard weak-coupling theory shows that the NVs relax towards equilibrium described by the same $T$ and $\mu$~\cite{PhysRevE.83.031111}. 
In contrast, the anisotropic dipole-dipole interaction between NV centers and the magnetic material breaks the conservation law of the total spin,  allowing the NV spins to change without an equal and opposite change in the environment's spin. 
This facilitates a genuine nonequilibrium steady state, where a continuous spin current is injected into the magnetic environment that tends to deplete the spin accumulation $\mu$. To stabilize a finite $\mu$ and maintain the spin grand-canonical ensemble of the environment, the magnet must be continuously pumped (irrespective of any intrinsic relaxation, i.e., Gilbert damping), which can be achieved by established spintronic methods such as microwave spin pumping~\cite{BRATAAS2006157}, spin-polarized transport~\cite{RevModPhys.76.323}, the spin Hall effect~\cite{RevModPhys.87.1213}, or the spin Seebeck effect~\cite{PhysRevB.81.214418}.

Once the environment is stabilized at a finite $\mu$, the nonequilibrium nature can be harnessed as a resource for steady-state entanglement generation between the NV centers.
Achieving this requires nonlocal dissipation between spatially separated NVs, which the magnetic environment can provide via magnon excitations. 
In the weak-coupling regime, the dissipative rates are determined by the environment fluctuations at the NV transition frequency.
The corresponding resonant magnon wavelength sets the length scale of the nonlocal dissipation. By tuning the NV splitting or the magnon dispersion, one can reach a regime where this length scale is comparable to or larger than the NV separation enabling correlated decay and excitation processes.
This nonlocal dissipation is most effective at low temperatures, where there is a pronounced imbalance between spontaneous emission and absorption processes, and the quantum-noise-dominated dynamics can stabilize entangled steady states.

While the NV-magnet setup is an ideal candidate to realize such a scheme, this concept is not restricted to spin degrees of freedom: any conserved quantity, such as particle number, or a topological invariant, can form the basis of an analogous scheme, provided the environment can be prepared with a finite chemical potential for that quantity and we have the means to break the corresponding conservation law.

\begin{figure}[t!]
    \centering
    \includegraphics[width=0.48\textwidth]{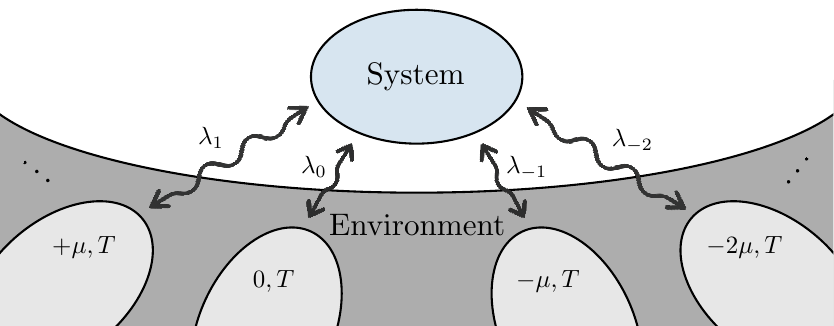}
    \caption{With a symmetry-breaking interaction, a pumped environment with chemical potential $\mu$ can be decomposed into multiple effective reservoirs with distinct chemical potentials $n\mu$ with $n\in \mathbb{Z}$.} 
    \label{fig:2}
\end{figure}

\textit{Effective multiple reservoirs}|We now show, in the weak-coupling regime, how an interaction that breaks a conservation law, together with an environment prepared in a grand canonical ensemble, effectively mimics a nonequilibrium setup in which the system couples to multiple reservoirs with different chemical potentials, see Fig.~\ref{fig:2}. Consider a system of $N$ independent qubits of energy $\Delta>0$, described by a Hamiltonian $H_S=\frac{\Delta}{2}\sum_i^N\sigma^z_i$, coupled to an environment $H_E$.
The environment is in a \gge of the form $\bar{\rho}_E=e^{-(H_{E}-\mu Q)/k_\text{B}T}/Z$ with $Z=\tr [e^{-(H_\text{E}-\mu Q)/k_\text{B}T}]$, where $Q=Q^\dagger$ is conserved in the environment, $\left[ H_{E},Q\right]=0$.

Generically, the coupling can be decomposed into terms distinguished by how they change the conserved quantity $Q$: 
\(\label{eq:H_I_gen}
H_I=\sum_n \lambda_n \sigma_i^- B_n(\vb{r}_i) + \mathrm{H.c.},
\)
where $\sigma^\pm_i = (\sigma^x_i\pm i\sigma^y_i)/2$ are Pauli operators on qubit $i$ at position $\vb{r}_i$, and the environment operators $B_n$ change the quantity $Q$ by $n$:
\(
[Q, B_n ]= n B_n.
\)
Note that $B^\dagger_n$ corresponds to a change of $Q$ by $-n$. When the sum contains only a single term (with coupling parameter $\lambda_n=\lambda \delta_{nm}$), the total quantity $Q_\text{tot}=m \sigma_z+Q$ remains conserved. Addition of at least one term with $n\neq m$ breaks the conservation law.

Each term in Eq.~\eqref{eq:H_I_gen} acts on the system as a distinct effective reservoir with chemical potential $n\mu$. To see this, note that under the Born-Markov and secular approximation for weak coupling~\cite{Breuer}, the system's emission and absorption processes are determined by the environment fluctuations~\cite{sm},
\begin{align}\label{eq:envfluct}
C^+_{n}(\omega,r_{ij})&= \int^\infty_{-\infty} \mathrm{d} t \ e^{i\omega t}  \ev{B^\dagger_{n}(\vb{r}_i,t)B_{n}(\vb{r}_j)},\\
C^-_{n}(\omega,r_{ij})&= \int^\infty_{-\infty} \mathrm{d} t \ e^{-i\omega t}  \ev{B_{n}(\vb{r}_i,t)B^\dagger_{n}(\vb{r}_j)},
\end{align}
respectively. 
Here, we assume a translationally invariant and isotropic environment, such that the correlations are real, $C^\pm_{n}(\omega,r_{ij})\in\mathbb{R}$, with $r_{ij}=\vert \vb{r}_i -\vb{r}_j \vert$ denoting the relative distance between sites. 
The $Q$ conservation law requires correlators between $B_n$ and $B_m^\dagger$  to be zero for $n\neq m$~\cite{sm}. The correlations $C^+_n(\omega,r)$ and $C^-_n(\omega,r)$ physically describe a process where the environment gains and loses energy $\hbar\omega$ and quantity $n$, respectively. 
In addition, they fulfill a distinct generalized Kubo-Martin-Schwinger~(KMS) condition dependent on $n$ (setting $\hbar=1$ from now on):
\begin{align}\label{eq:nKMS}
    \frac{C^-_{n}(\omega,r)}{C^+_{n}(\omega,r)}=e^{-(\omega-n \mu)/k_\text{B}T}.
\end{align}
If the interaction $H_I$ contains only a single term, $\lambda_n=\lambda \delta_{nm}$, the system will thermalize to a Gibbs ensemble at temperature $T$ and chemical potential $m\mu$~\cite{PhysRevE.83.031111}.
If the interaction $H_I$ contains multiple terms, each $n$ channel tends to drive the system toward a different generalized equilibrium characterized by $n\mu$, see Fig.~\ref{fig:2}. Hence, for finite $\mu$ the system generically approaches a nonequilibrium steady state. When $\mu = 0$, all channels correspond to the same thermal distribution at $T$, and the system thermalizes.

\textit{Nonequilibrium dynamics}|In the nonequilibrium setting described above, we now analyze under what conditions nonlocal dissipation can generate steady-state entanglement between two qubits.
We determine the steady-state entanglement of a two-qubit system by considering the dynamics of the state $\rho_S$ under the Born-Markov secular approximation $\dot{\rho}_S=-i[H_S,\rho_S] + \sum_{ij}\mathcal{L}_{ij}[\rho_S]$. The Lindblad superoperator $\mathcal{L}_{ij}[\rho_S]=\mathcal{L}^e_{ij}[\rho_S]+\mathcal{L}^a_{ij}[\rho_S]$ describes the dissipation,
\(\al{
\mathcal{L}^e_{ij} [\rho_S] =  \ &\Gamma_e(r_{ij}) \Big( \sigma^-_i \rho_S \sigma^+_j - \frac{1}{2}\{\sigma^+_j \sigma^-_i, \rho_S\} \Big), \\
\mathcal{L}^a_{ij} [\rho_S] =   \ &\Gamma_a(r_{ij}) \Big( \sigma^+_i \rho_S \sigma^-_j - \frac{1}{2}\{\sigma^-_j \sigma^+_i, \rho_S\} \Big),
}\) 
where the nonlocal emission and absorption rates are related to the environment correlations from Eq.~\eqref{eq:envfluct} by 
\(\al{\label{eq:rate_equation}
&\Gamma_e(r)=\sum_n\abs{\lambda_n}^2 C^+_n(\Delta,r), \\ 
&\Gamma_a(r)=\sum_n\abs{\lambda_n}^2 C^-_n(\Delta,r).
}\)
While we consider the case of $N=2$ qubits, these equations are valid for arbitrary numbers of qubits with pairwise separations $r_{ij}$. 

\begin{figure}[t!]
    \centering
    \includegraphics[width=0.47\textwidth]{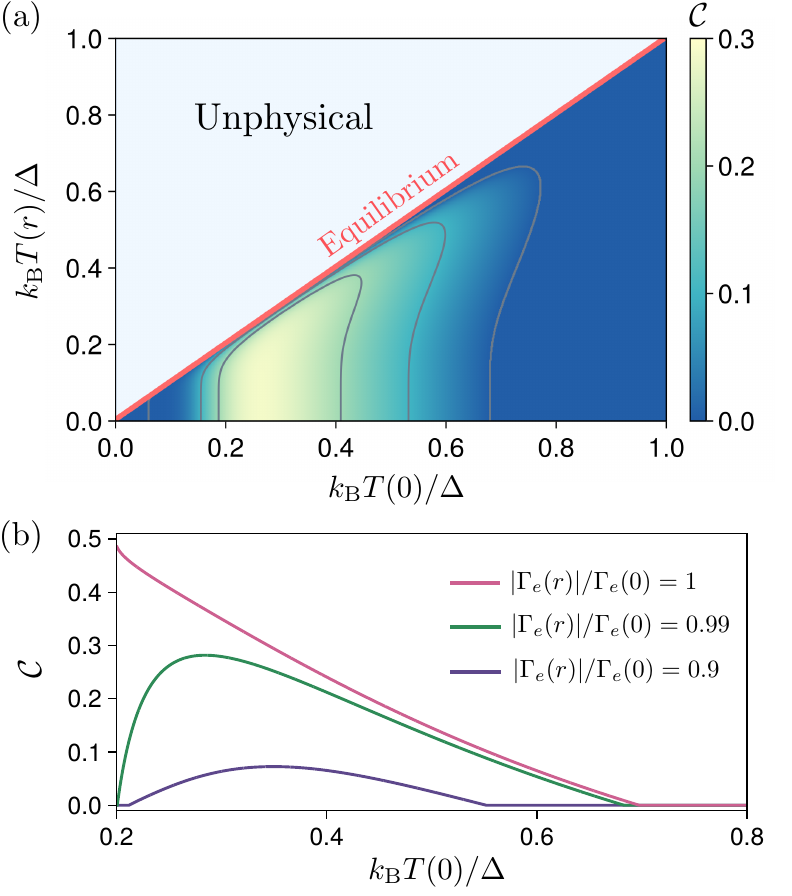}
    \caption{Steady‐state concurrence $\mathcal{C}$ of a two-qubit system as a function of (a) $T(0)$ and $T(r)$, with $\abs{\Gamma_e(r)}/\Gamma_e(0)=0.99$ exemplifying strong nonlocal dissipation, where unphysical region corresponds to temperatures violating the positivity of the Lindblad evolution~\cite{sm}, and (b) local temperature $k_\text{B}T(0)/\Delta$ for fixed nonlocal temperature $k_\text{B}T(r)/\Delta = 0.2$. For $k_\text{B}T(0)/\Delta$ below this value, the system approaches the unphysical region, whose precise extent depends on $\abs{\Gamma_e(r)}/\Gamma_e(0)$, and no entanglement is observed~\cite{sm}.}
    \label{fig:3}
\end{figure}

We choose $\Gamma_e(0)$ to set the overall timescale of the dynamics. 
The steady state is then determined by three parameters: the relative strength of the nonlocal emission process $\abs{\Gamma_{e}(r)}/\Gamma_{e}(0)$, and two \textit{effective temperatures} $T(0)$ and $T(r)$, which characterize the local and nonlocal dissipation, respectively~\footnote{ In addition, the signs of $\Gamma_{e}(r)$ and $\Gamma_{a}(r)$ are relevant, as the nonlocal rates may carry a nonzero phase of $\pi$. This phase has no qualitative effect on entanglement generation for two qubits and is therefore omitted in the general discussion. In a multi‐qubit system, however, relative phases between nonlocal processes are expected to play a more prominent role. Furthermore, complete positivity of the Lindblad equation imposes two constraints on the rates~\cite{sm}: $0\leq\abs{\Gamma_{e}(r)}/\Gamma_{e}(0)\leq1$ and $0\leq\abs{\Gamma_{a}(r)}/\Gamma_{a}(0)\leq1$.}. This notion of effective temperature $T$ is defined by 
\(\label{eq:T}
e^{-\Delta/k_\text{B}T(r)}\equiv|\Gamma_{a}(r)/\Gamma_{e}(r)|.
\)
When $T(0)=T(r)$, the system will relax to a thermal state $\bar{\rho}_S=e^{-H_S/k_\text{B}T(0)}/Z$ that satisfies the detailed balance condition $\mathcal{L}^e_{ij} [\bar{\rho}_S]+\mathcal{L}^a_{ji} [\bar{\rho}_S]=0$ both for local ($i=j$) and nonlocal ($i\neq j$) processes~\cite{sm}. Hence, to avoid thermalization, we require $T(0)\neq T(r)$.

\textit{Entanglement generation}|We now quantify the resulting steady-state entanglement using the concurrence $\mathcal{C}$ \cite{PhysRevLett.80.2245}, which ranges from 0 for separable states to 1 for maximally entangled Bell pairs. Fig.~\ref{fig:3}a shows that the entanglement is stabilized at an intermediate local temperature $T(0)\sim\Delta/k_\text{B}$ and can persist when the nonlocal temperature is lower, $\abs{T(r)}< \abs{T(0)}$. While entanglement is stabilized at a finite local temperature, it remains a fundamentally low-temperature phenomenon ($\abs{T(0)}\lesssim\Delta/k_\text{B}$), where quantum correlations dominate thermal noise. The magnitude of the concurrence is primarily controlled by the relative strength of the nonlocal process, as seen in Fig.~\ref{fig:3}b. Significant entanglement is obtained for $\abs{\Gamma_e(r)}/\Gamma_e(0) > 0.9$.

A pumped environment can also give rise to negative temperatures $T(0)<0$ and $T(r)<0$, analogous to population inversion in lasing systems \cite{svelto_principles_of_lasers}. We find that the entangled steady states also arise in this regime, particularly when the absorption process is sufficiently nonlocal, parameterized by $\abs{\Gamma_a(r)}/\Gamma_a(0)\approx1$~\cite{sm}. Due to the symmetry between emission and absorption, the analysis for $T(0),T(r)>0$ carries over to $T(0),T(r)<0$ upon replacing $e\mapsto a$. In both cases, the entangled steady states are unique mixed states with a finite fraction in the singlet state $(\ket{\uparrow\downarrow}-\ket{\downarrow\uparrow})/\sqrt{2}$. By contrast, no entanglement is observed when $T(0)$ and $T(r)$ have opposite signs, suggesting that a consistent dominance of either emission or absorption, both locally and nonlocally, is necessary. When the dominant process is sufficiently nonlocal, the system can reach a steady state with appreciable concurrence.

\begin{figure}[t!]
    \centering
    \includegraphics[width=0.45\textwidth]{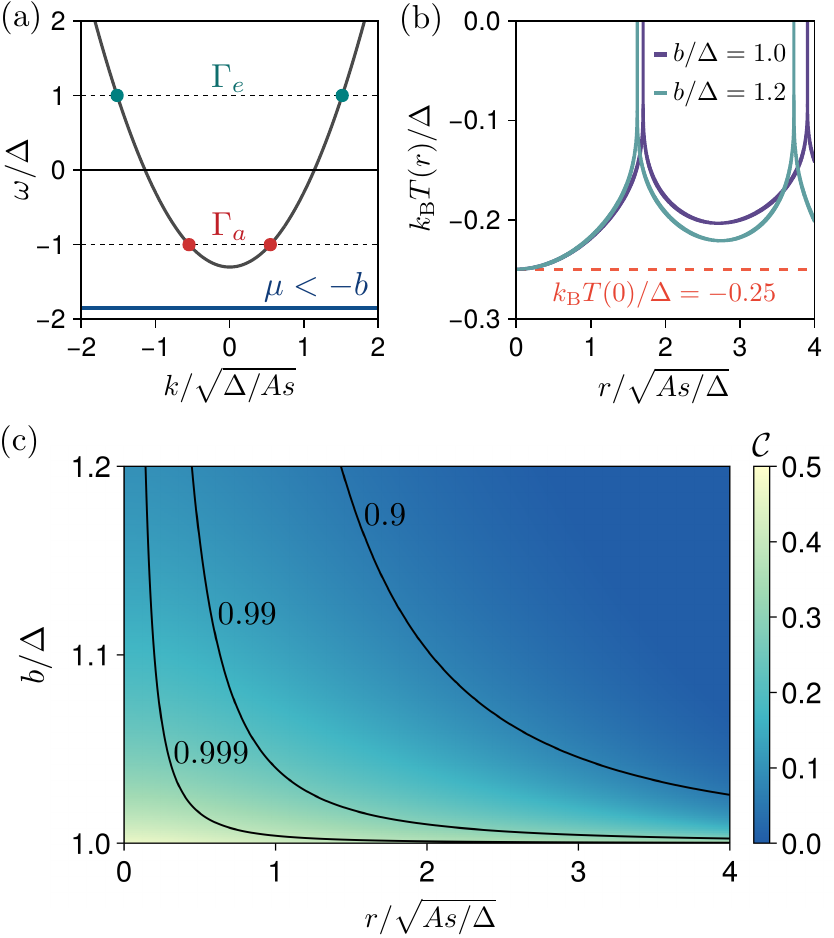}
    \caption{(a) Dispersion relation of magnetic excitations of the \textit{inverted} magnet. The state is stabilized by a spin accumulation $\mu<-b$, set by the external field. The crossings at $\omega=\pm\Delta$ indicate excitations associated with the emission and absorption processes of the system. 
    (b) Nonlocal temperature $k_\text{B}T(r)/\Delta$ as a function of qubit separation $r/\sqrt{As/\Delta}$ for $b/\Delta=1.0$ and $1.2$. We choose the coupling ratio to be $\abs{\lambda_1/\lambda_{-1}}=0.135$, which sets local temperature at $k_{B}T(0)/\Delta=-0.25$. The zeros of $T(r)$ corresponds to the zeros of the Bessel function.
    (c) Steady-state concurrence $\mathcal{C}$ as a function of $b/\Delta$ and $r/\sqrt{As/\Delta}$. The same coupling ratio is used as in (b). The condition $b/\Delta \geq 1$ ensures that both emission and absorption processes are present. Contour lines of the relative strength of the nonlocal absorption, $\abs{\Gamma_a(r)}/\Gamma_a(0) = 0.999$, $0.99$, and $0.9$, are overlaid.}
    \label{fig:4}
\end{figure}

\textit{Model based on spintronics}|To demonstrate our scheme, we consider a quantum system composed of two NV centers functioning as spin qubits. The spin-pumped environment is modeled as a two-dimensional (2D) inverted Heisenberg magnet~\cite{10.1063/5.0151652}:
\( 
{H}_E=\int \mathrm{d}^2r\left[\frac{A}{2}\left(\nabla \vb{s}\right)^2 - bs_z\right], 
\)
where $\vb s(\vb r)$ is the spin density operator, $A$ is the spin stiffness, which favors parallel alignment of spins, and $b>0$ is the applied field. The magnet is pumped and described by the spin grand-canonical ensemble $\bar\rho_E=e^{-\left( H_E -\mu S_z\right)/k_\text{B}T}/Z$, where $\mu$ is the spin accumulation and $S_z=\int\mathrm{d}^2r\,{s}_z$ is the total spin-$z$ component~\cite{PhysRevB.66.224403}. 
To illustrate our scheme, we consider a strong spin accumulation $\mu<-b$ and temperature $T\rightarrow 0$, so that the environment is in the inverted pure state $\ket{\downarrow\downarrow\downarrow\cdots}$. 
The inverted state hosts magnonic excitations with positive spin only. However, they can carry both positive and negative energies, see Fig.~\ref{fig:4}(a).

Furthermore, we consider a minimal interaction Hamiltonian $H_I$ that contains both a spin-conserving component $\lambda_1$ and a spin-nonconserving component $\lambda_{-1}$~\footnote{Alternatively, the interaction in Eq.~\eqref{eq:H_I} could be realized via the stronger exchange coupling between spin-qubits other than NVs, e.g. quantum dots, and the magnet, provided the system's quantization axis is tilted relative to the environment magnetization. In this case, the global spin-$z$ conservation is broken by the axes misalignment, and the coupling constants $\lambda_{1}$ and $\lambda_{-1}$ are set by the relative angle between the orientations~\cite{sm}.}:
\(\label{eq:H_I}
H_{I}
= \sum_i \lambda_1\sigma^-_i s_+(\vb{r}_i) + \lambda_{-1}\sigma^-_i s_-(\vb{r}_i) + \mathrm{H.c.}.
\)
Here, the NVs with Pauli operators $\sigma^\pm_i$ at position $\vb{r}_i$ are locally coupled to the environment's spin density field $s_\pm(\vb r_i) = [s_x(\vb{r}_i)\pm is_y(\vb{r}_i)]/2$.
This serves as a toy model that mimics the anisotropy of the dipole coupling between NVs and a magnetic material. 
Using our insights from Fig.~\ref{fig:2}, the breaking of spin conservation by $H_I$ allows us to view the pumped magnet as two effective reservoirs of distinct chemical potentials $+\mu$ and $-\mu$. 
For $b \ge \Delta$, this leads to both emission and absorption processes with rates~\cite{sm},
\(\al{
&\Gamma_{e}(r) = 4\pi s\abs{\lambda_1}^2 g_{2D} J_0[k(\Delta)r],\\
&\Gamma_{a}(r) = 4\pi s\abs{\lambda_{-1}}^2 g_\text{2D} J_0[k(-\Delta)r],
}
\)
where $J_0(x)$ is the zeroth-order Bessel function, $k(\omega) = \sqrt{(\omega+b)/As}$ is the magnon wavevector for saturated spin density $s$, and $g_\text{2D} = \frac{1}{4\pi As}$ is the 2D density of states, with saturated spin density $s$. 
These rates describe the processes in which the qubit emits~(i.e. $\Gamma_e$) energy by creating a magnon of energy $\Delta$ and wave vector $k(\Delta)$, or absorbs~(i.e. $\Gamma_a$) energy by creating a magnon of negative energy $-\Delta$ (also referred to as an antimagnon~\cite{10.1063/5.0151652}) with wave vector $k(-\Delta)$. 
In both cases the environment's total spin increases and the spin accumulation $\mu$ is reduced, so maintaining the inverted ensemble requires continuous pumping.

As shown in Fig.~\ref{fig:3}b, to generate entanglement, the nonlocal dissipation must remain appreciable at the separation between the two NVs.
In that figure, a positive local temperature is considered, and strong nonlocal emission $\abs{\Gamma_e(r)}\approx \Gamma_e(0)$ is required.
For the inverted magnet, however, the \textit{absorption} channel is naturally more long-ranged because $k(-\Delta)<k(\Delta)$, so that the spatial dependence of $\Gamma_a(r)$ decays more slowly than that of $\Gamma_e(r)$. We therefore focus on the regime in which absorption is the dominant process and strong entanglement corresponds to $\abs{\Gamma_a(r)} \approx \Gamma_a(0)$.

We fix the ratio of couplings to be $\abs{\lambda_1/\lambda_{-1}} = 0.135$, which yields a negative local temperature, $T(0)=\Delta/(2k_B\ln\left|\lambda_1/\lambda_{-1}\right|)<0$, and compute the steady-state concurrence as a function of $b/\Delta$ and $r/\sqrt{As/\Delta}$, shown in Fig.~\ref{fig:4}c. Entanglement is found for NV separation $r\lesssim 2\sqrt{As/\Delta}$ and for external fields in the range $\Delta\leq b\lesssim 1.1\Delta$. When $b=\Delta$, the wavevector $k(\Delta)$ vanishes and $\abs{\Gamma_a(r)}/\Gamma_a(0)=1$ for all distances, allowing entanglement to persist even at large separations. For larger $b$, the ratio $\abs{\Gamma_a(r)}/\Gamma_a(0)$ decreases with distance and the concurrence is reduced. Overall, the entanglement is largely controlled by the relative strength of the nonlocal process $\abs{\Gamma_a(r)}/\Gamma_a(0)$, while it is only weakly sensitive to the nonlocal temperature, as illustrated in Fig.~\ref{fig:4}b.

Reducing the qubit separation $r$ increases the steady-state concurrence, which is maximized at $r = 0$. This enhancement, however, comes at the cost of a reduced convergence rate~\cite{sm}, in line with the general trade-off between steady-state entanglement and relaxation time discussed in Ref.~\cite{PRXQuantum.5.040305}. In particular, at $r=0$ the entangled steady state is approached only in the limit $t \to \infty$, so that the order of limits becomes essential: this state is obtained only if the $t \to \infty$ limit is taken before sending $r \to 0$.

\textit{Discussion}|We have demonstrated that steady-state entanglement can be stabilized by a spin-pumped environment with an anisotropic, spin-nonconserving coupling to spin qubits. 
The environment itself is described by a spin grand-canonical ensemble that satisfies the fluctuation-dissipation relation, Eq.~\eqref{eq:nKMS}, while the combination of a finite spin accumulation and the conservation-law-breaking interaction drives the qubits into a nonequilibrium steady state. In this regime, a two-qubit system can acquire steady-state entanglement at low temperatures.

In contrast to the environment-decoupled dark states, here, the system naturally evolves into the unique entangled steady state through local and nonlocal emission and absorption processes.
The correlated processes that support this behavior arise when the resonant magnon wavelength is larger than the NV separation, a regime that is reminiscent of superradiance and subradiance in optical systems~\cite{GROSS1982301} but mediated by long-wavelength magnons.

Small NV spacing may be possible in experiments coupling NVs and yttrium iron garnet thin  films~\cite{PhysRevB.111.064424,PhysRevApplied.15.034031,le2025widebandcovariancemagnetometrydiffraction}.
We expect a key obstacle an experiment might face is obtaining strong enough dissipative coupling to overcome intrinsic dephasing.
For magnetic dipole interactions, the coupling $\lambda_{-1}$ to long wavelength magnon decreases linearly with $k(-\Delta)$, which results in a tradeoff between the coupling strength $\propto k(-\Delta)$ and the spatial range of the dissipative channel $k(-\Delta)^{-1}$.
This constraint could be mitigated by using a softer magnet with larger density of states, which enhances the dissipative rates.

Our simple mechanism to controllably drive a system of qubits out of equilibrium may be relevant to other platforms and symmetries.
For example, impurity qubits coupled to an integrable quantum gas may naturally evolve into an entangled steady state~\cite{PhysRevLett.98.050405,Vasseur_2016}: the integrable gas relaxes to a generalized Gibbs ensemble with an extensive number of conservation laws, which are naturally broken by the coupling to the impurity qubits. 
A perhaps simpler solid-state platform is an antiferromagnet, where again a coupling that breaks a spin conservation law is necessary, but where the flexible tuning of the nonlocal coupling can arise from the two magnon branches of the antiferromagnet, obviating the need for an inverted state.

A natural direction for future work is to explore the many-body nonequilibrium dynamics generated by this mechanism. 
In color-center platforms, two-qubit quantum state tomography and entanglement verification using spin-dependent optical readout have already been demonstrated~\cite{PhysRevLett.130.090801,sciadv.1501015}, so the steady state we predict, including its finite singlet component, should in principle be experimentally accessible. 
Beyond verification, entanglement is a central resource for sensing~\cite{Rovny2025} and computational tasks, and the broader outlook of this approach lies in its potential application to scaled-up systems.
For example, a similar protocol may be able to stabilize spin-squeezed states between two or more ensembles of qubits.
On the theoretical side, understanding the structure of such nonequilibrium many-body steady states is a challenging problem. However, the recently developed `hidden time-reversal symmetry'~\cite{PRXQuantum.2.020336} may suggest strategies for devising scaled-up models that are analytically solvable.

\textit{Acknowledgments}|We thank Ji Zou and Jamir Marino for particularly valuable discussions, and Yanyan Zhu and Shu Zhang for additional helpful input. We also thank Aashish Clerk for insights on future directions.
This work is supported by the U.S. Department of Energy, Office of Basic Energy Sciences under Grant No. DE-SC0012190.

\bibliography{citations}

\end{document}



\def\im{\mrm{im}}
\def\tr{\mrm{tr}}
\def\Z{\mbb{Z}}
\def\R{\mbb{R}}
\def\C{\mbb{C}}
\def\p{\partial}

\renewcommand\({\begin{equation}}
\renewcommand\){\end{equation}}

\newcommand{\al}[1]{\begin{aligned}#1\end{aligned}}
\newcommand{\eq}[1]{\begin{equation}#1\end{equation}}

\newcommand{\ph}[1]{\phantom{#1}}
\newcommand{\mbb}[1]{\mathbb{#1}}
\newcommand{\mrm}[1]{\mathrm{#1}}
\newcommand{\mcal}[1]{\mathcal{#1}}
\newcommand{\mbf}[1]{\mathbf{#1}}

\newcommand{\D}[2]{\frac{d#1}{d#2}}
\newcommand{\pd}[2]{\frac{\p#1}{\p#2}}

\title{Supplementary Material for Breaking a conservation law enables steady-state entanglement out of equilibrium}
\author{Vince Hou}
\author{Eric Kleinherbers}
\author{Shane P. Kelly}
\author{Yaroslav Tserkovnyak}
\affiliation{Department of Physics and Astronomy and Bhaumik Institute for Theoretical Physics, University of California, Los Angeles, California 90095, USA}
\date{\today}
\maketitle

In this Supplemental Material, we provide (i) a derivation of the generalized KMS condition, (ii) a proof of complete positivity of the Lindblad master equation, from both a stationarity perspective and a thermodynamic perspective, (iii) a discussion on the numerical results of steady state concurrence and the convergence rate, (iv) a proof of the detailed balance condition satisfied by the thermal Gibbs state, (v) a discussion on how to break conservation law with an exchange coupling between the system and the environment, and (vi) a derivation of the system emission and absorption rates given by the inverted magnet environment.

\section{Generalized Kubo-Martin-Schwinger Condition}

We elaborate the generalized Kubo-Martin-Schwinger (KMS) relation of an environment described by a grand-canonical ensemble. The environment has a chemical potential $\mu$ controlled by the population of a conserved quantity $Q$, $[H_E, Q] = 0$, where $H_E$ is the environment Hamiltonian. The environment state is given by a grand-canonical ensemble,
\(
\bar\rho_E = \frac{e^{-\beta(H_E-\mu Q)}}{\operatorname{Tr} [e^{-\beta(H_E-\mu Q)}]}.
\)
Here $\beta = 1/k_\mathrm{B} T$, where $T$ is the environment temperature. We now show that the correlation functions $C^\pm_{nm}(\omega,r_{ij})$ defined in the main article, 
\(\label{eq:envfluct}
C^+_{nm}(\omega,r_{ij})= \int^\infty_{-\infty} \mathrm{d} t \ e^{i\omega t}  \ev{B^\dagger_{n}(\vb{r}_i,t)B_{m}(\vb{r}_j)},
\quad
C^-_{nm}(\omega,r_{ij})= \int^\infty_{-\infty} \mathrm{d} t \ e^{-i\omega t}  \ev{B_{n}(\vb{r}_i,t)B^\dagger_{m}(\vb{r}_j)},
\)
satisfy
\(\label{eq:KMS_C}
C^+_{nm}(\omega,r_{ij})=\delta_{nm}e^{\beta(\omega-n\mu)}C^-_{mn}(-\omega,r_{ji}).
\)
Suppose we have environment operators $B_n(\vb r_i)$ and $B_m(\vb r_j)$ that satisfy the commutation relations
\(
[Q, B_n] = nB_n, \quad [Q, B^\dagger_n]=-nB^\dagger_n,
\)
where $n$ indicates the corresponding change in $Q$ and we drop the position dependence for brevity.  
We start by evaluating the correlation functions with respect to $\bar\rho_E$ 
\(\al{
\langle B^\dagger_n(t)B_m\rangle = \operatorname{Tr} [e^{-\beta(H_E-\mu Q)} e^{iH_E t}B^\dagger_n e^{-iH_E t} B_m]/Z,
}\)
where $Z=\operatorname{Tr} [e^{-\beta(H_E-\mu Q)}]$. First, we insert $I=e^{-\beta H_E} e^{\beta H_E}$ after $B_n^\dagger$ to obtain,
\(
\langle B^\dagger_n(t)B_m\rangle = \operatorname{Tr} [e^{-\beta H_E} e^{\beta\mu Q} e^{iH_E t}B^\dagger_n e^{-\beta H_E} e^{\beta H_E} e^{-iH_E t} B_m]/Z.
\)
Using the fact that trace is invariant under cyclic permutations and $[H_E, Q] = 0$,
\(
\langle B_n^\dagger(t)B_m\rangle 
= \operatorname{Tr} [e^{-\beta H_E} B_m e^{\beta\mu Q} B^\dagger_n(t+i\beta)]/Z. \label{eq:KMSstep}
\)
We multiply the operator $B^\dagger_n$ by $I=e^{-\beta\mu Q} e^{\beta\mu Q}$ from the right and perform cyclic permutations again,
\(\al{
\langle B_n^\dagger(t)B_m\rangle 
&= \operatorname{Tr} [e^{-\beta H_E} B_m e^{\beta\mu Q} B^\dagger_n(t+i\beta) e^{-\beta\mu Q} e^{\beta\mu Q}]/Z\\
&= \operatorname{Tr} [e^{-\beta(H-\mu Q)} B_m e^{\beta\mu Q} B^\dagger_n(t+i\beta) e^{-\beta\mu Q}]/Z.
}\)
From the commutation relation $[Q, B^\dagger_n] = -nB^\dagger_n$, we conclude that $e^{\beta\mu Q} B^\dagger_n e^{-\beta\mu Q} = e^{-\beta\mu n}B^\dagger_n$, such that,
\(\label{eq:KMS_1}
\langle B_n^\dagger(t)B_m\rangle = e^{-\beta\mu n} \langle B_mB^\dagger_n(t+i\beta)\rangle.
\)
Similarly, one can use Eq.~\eqref{eq:KMSstep} and multiply the operator $B_m$ by $I=e^{-\beta\mu Q} e^{\beta\mu Q}$ from the left to derive an equivalent form,
\(\label{eq:KMS_2}
\langle B^\dagger_n(t)B_m\rangle = e^{-\beta\mu m} \langle B_mB^\dagger_n(t+i\beta)\rangle.
\)
The only way for Eq.~\eqref{eq:KMS_1} and Eq.~\eqref{eq:KMS_2} to be equal, is by setting $\langle B^\dagger_n(t) B_m \rangle=0$ for $n \neq m$.
As a result, only diagonal correlation functions are nonzero, i.e.,
\(\label{smeq:correlation_nm}
\langle B^\dagger_n(t)B_m\rangle = \delta_{nm}e^{-\beta\mu n} \langle B_mB^\dagger_n(t+i\beta)\rangle.
\)
The Fourier transform of this generalized KMS condition, after shifting $t$ integration to $t-i\beta$, has expression
\(
\int^\infty_{-\infty} \mathrm{d}t \ e^{i\omega t}\langle B^\dagger_{n}(\vb{r}_i,t)B_m(\vb{r}_j) \rangle 
= \delta_{nm}e^{\beta(\omega-n\mu)}\int^\infty_{-\infty} \mathrm{d}t \ e^{i\omega t}\langle B_m(\vb{r}_j)B^\dagger_n(\vb{r}_i,t) \rangle, 
\)
where we include the position dependence $\vb r_i$ of the operators. Hence we have proven Eq. \eqref{eq:KMS_C}.

\section{Derivation of Lindblad Master Equation}
In this section we summarize the standard weak-coupling derivation of the Lindblad equation used in the main text, following the Born-Markov secular procedure (see Ref.~\cite{Breuer}). We consider a system with Hamiltonian $H_S=\frac{\Delta}{2}\sum_i^N\sigma^z_i$ coupled to an environment via the interaction $H_I = \sum_{i,n} \lambda_n \sigma_i^- B_n(\mathbf r_i) + \mathrm{H.c.}$, where $B_n(\mathbf r_i)$ are environment operators that change the conserved quantity $Q$ by $n$.

Treating $H_I$ perturbatively to second order (Born approximation) and assuming a stationary environment state $\bar\rho_E$, we obtain the Born-Markov master equation for the reduced density matrix $\rho_S$ in the Schr\"odinger picture,
\(\label{eq:Born-Markov}  
\dot{\rho}_S=-i[H_{S},\rho_S]-\int_0^\infty \mathrm{d}\tau \; \operatorname{Tr}_E \Big\{  \Big[ H_I, [\tilde{H}_I(-\tau),\rho_S\otimes\bar{\rho}_E   ]   \Big]   \Big\},
\)
where the notation $\tilde{O}(t)=e^{i(H_S+H_E)t}Oe^{-i(H_S+H_E)t}$ indicates an operator in the interaction picture of $H_S+H_E$. 
Inserting the interaction $H_I$ and decomposing the system operators into eigenoperators of $H_S$ with Bohr frequencies $\pm\Delta$, the time integrals pick out the environment correlation functions at these frequencies, i.e., the system operators acquire time dependence
$\sigma_i^\pm(t) = e^{\pm i \Delta t} \sigma_i^\pm$. The Markov approximation amounts to extending the upper limit of the $\tau$-integration to infinity, and the secular approximation discards rapidly oscillating terms, keeping only contributions whose net phase is time-independent. This leaves dissipative terms resonant at the Bohr frequencies $\pm \Delta$, with rates determined by the environment correlation functions $C_n^\pm(\Delta,r)$, defined in Eq.~(\ref{eq:envfluct}).

The terms can be grouped into two contributions: (i) an effective Hamiltonian $H_{\mathrm{LS}}$, commonly referred to as the Lamb-shift Hamiltonian, and (ii) a dissipative part $\mathcal{L}[\rho_S]$, often termed the Lindbladian:
\(\label{eq:full_lindblad}  
\dot{\rho}_S = -i[H_S+H_\text{LS}, \rho_S] + \mathcal{L}[\rho_S].
\)
The effective Hamiltonian contains both induced coherence couplings and an energy level shift, (with constant terms discarded),
\(
H_\text{LS} = \sum_{i\neq j}(\mathcal{J}_{ij}\sigma_i^+\sigma_j^- + \mathrm{H.c.})+\sum_i\frac{\delta_i}{2}\sigma^z_i.
\)
The induced coherence has coupling strength:
\(
\mathcal{J}_{ij} = \sum_n\frac{\lambda_n^2}{2}\mathfrak{Re}G^R[\Delta;B_n(\vb r_j),B^\dagger_{n}(\vb r_i)].
\)
Here we adopt the definition for retarded Green's function $G^R(\omega;A,B) = -i\int_0^\infty \mathrm{d}t\ e^{i\omega t}\langle[A(t),B]\rangle$ and advanced Green's function defined by $G^A(\omega;A,B) = -i\int_0^\infty \mathrm{d}t\ e^{i\omega t}\langle[A,B(t)]\rangle$,
and denote the `real' part of the retarded Green's function as $\mathfrak{Re}G^R(\omega;A,B)\equiv[G^R(\omega;A,B)+G^A(\omega;A,B)]/2$. The energy shift term has magnitude,
\(
\delta_i = \sum_n\abs{\lambda_n}^2\,\text{Im}\Big[\int_0^\infty\mathrm{d}\tau \ e^{-i\Delta\tau}\langle\{B^\dagger_{n}(\vb{r}_i,\tau), B_{n}(\vb r_i)\}\rangle\Big].
\)

The Lindblad superoperator is given by
\(\label{eq:Lindblad_superoperator}
\mathcal{L}[\rho_S] =\sum_{ij} \Gamma_e(r_{ij})\Big(\sigma^-_j\rho_S\sigma^+_i-\frac{1}{2}\{\sigma^+_i\sigma^-_j, \rho_S\}\Big)
+\Gamma_a(r_{ij})\Big(\sigma^+_j\rho_S\sigma^-_i-\frac{1}{2}\{\sigma^-_i\sigma^+_j, \rho_S\}\Big),
\)
where the emission and absorption rates are given by the environment correlations,
\(
\Gamma_e(r)=\sum_n\abs{\lambda_n}^2 C^+_n(\Delta,r), \quad
\Gamma_a(r)=\sum_n\abs{\lambda_n}^2 C^-_n(\Delta,r).
\)
Here we have assumed that the environment is translationally invariant and isotropic, which guarantees $C^\pm_n(\Delta,r)\in\mathbb{R}$. Therefore, the relative phase between local and nonlocal rates are constrained to two possible values, $\pm 1$, i.e., $\Gamma_e(r)=\pm\abs{\Gamma_e(r)}$ and $\Gamma_a(r)=\pm\abs{\Gamma_a(r)}$.

For a two-qubit system, the Lindblad superoperator can be brought to its diagonal form, 
\(\label{eq:Lindblad_diagonal}
\mathcal{L}[\rho_S] = \sum_{\alpha}\gamma_\alpha \Big( L_\alpha \rho L^\dagger_\alpha - \frac{1}{2}\{L^\dagger_\alpha L_\alpha, \rho_S\} \Big),
\) 
with four quantum-jump operators $L_\alpha$ and their corresponding damping rates $\gamma_\alpha$,
\(\label{eq:jump_operator}\al{
&\gamma_1 = \sqrt{\Gamma_a(0)+\abs{\Gamma_a(r)}}, \quad 
L_1 = \sigma^+_1+\mathrm{sgn}\big[\Gamma_a(r)\big]\sigma^+_2, \quad
\gamma_2 = \sqrt{\Gamma_a(0)-\abs{\Gamma_a(r)}}, \quad 
L_2 = \sigma^+_1-\mathrm{sgn}\big[\Gamma_a(r)\big]\sigma^+_2, \\
&\gamma_3 = \sqrt{\Gamma_e(0)+\abs{\Gamma_e(r)}}, \quad 
L_3 = \sigma^-_1+\mathrm{sgn}\big[\Gamma_e(r)\big]\sigma^-_2, \quad
\gamma_4 = \sqrt{\Gamma_e(0)-\abs{\Gamma_e(r)}}, \quad 
L_4 = \sigma^-_1-\mathrm{sgn}\big[\Gamma_e(r)\big]\sigma^-_2. 
}\)
This is the Lindblad generator used in the main text: for $N=2$ we specialize to two qubits with the local and nonlocal emission and absorption rates defined there.

\section{Constraints on system rates}
The coefficient matrix of the Lindblad equation must be a positive semidefinite (PSD) matrix to ensure trace-preserving and completely positive dynamics. In the present setting, this requirement imposes $\Gamma_e(0)\geq\abs{\Gamma_e(r)}$ and $\Gamma_a(0)\geq\abs{\Gamma_a(r)}$ for the emission and absorption rates, respectively. Below, we present derivations of these constraints from two perspectives. 

\subsection{Stationary environment}
We first prove that when the environment is in a general \textit{stationary} state $\bar\rho_E$, and not necessarily thermal, the PSD condition is still guaranteed. Define operator:
\begin{align}\label{eq:suppO}
O(t) = \sum_{i=1}^N \lambda_i B_n(\vb r_i, t),
\end{align}
for arbitrary complex coefficients $\lambda_i$ with index $i\in\{1,\ldots,N\}$ denoting the qubits at position $\vb r_i$. Then, we define the two-point correlation function $K(t_1,t_2)=\langle O^\dagger(t_1) O(t_2) \rangle$ and apply the Fourier transform:
\(
S(\omega)=\int_{-\infty}^{\infty} \mathrm{d}t\, e^{i\omega t} K(t,0)  = \sum_{i,j} \lambda_i^* \lambda_j C^+_n(\omega, r_{ij}),
\)
where in the second step we inserted Eq.~\eqref{eq:suppO} to identify the fluctuations $C^+_n(\omega, r_{ij})$.
To prove $C^+_n(\omega, r_{ij})$ is PSD, it remains to show that 
\(
S(\omega)\geq 0.
\)
To do so, we first show the positivity of the kernel $K(t_1,t_2)$ :
\begin{align}
\int_{-\infty}^{\infty} \mathrm{d}t_1\int_{-\infty}^{\infty} \mathrm{d}t_2\, c^*(t_1) K(t_1,t_2) c(t_2) = \operatorname{Tr}[\bar\rho_E\, \Phi^\dagger \Phi]\geq 0, \label{eq:supppos}
\end{align}
where we have defined $\Phi=\int_{-\infty}^{\infty} \mathrm{d}t\, c(t)O(t)$. Since the density matrix $\bar\rho_E$ is a positive operator, and $\Phi^\dagger \Phi \geq 0$, the trace of their product is non-negative.
Next, we rewrite the integral as
\begin{align}
&\phantom{{}=}\int_{-\infty}^{\infty} \mathrm{d}t_1\int_{-\infty}^{\infty} \mathrm{d}t_2\, c^*(t_1) K(t_1,t_2) c(t_2)\\
&=\int_{-\infty}^{\infty} \mathrm{d}\tau  \,  K(\tau,0) \int_{-\infty}^{\infty}\mathrm{d}t \,c^*(t)c(t-\tau)\\
&=\int_{-\infty}^{\infty} \frac{\mathrm{d}\omega}{2\pi}  \,  S(\omega) \abs{c(\omega)}^2\geq 0,
\end{align}
where in the first step we switch to variables $\tau=t_1-t_2$ and $t=t_1$ and use stationarity $K(t_1,t_2)=K(t_1-t_2,0)$. In the second step, we identified the convolution and inserted the Fourier transforms. According to Eq.~\eqref{eq:supppos}, positivity is fulfilled for arbitrary functions $c(t)$, and, thus, for arbitrary Fourier transforms $c(\omega)$. Therefore,  we infer that $S(\omega)\geq 0 $ for all $\omega$.

Likewise, $C^-_{n}(\omega, r_{ij})$ can be shown to be PSD in the same spirit. 

\subsection{Second law of thermodynamics}
We now demonstrate, from a thermodynamic perspective, that for a generalized thermal environment state $\bar\rho_E = e^{-\beta(H_E - \mu Q)}/Z$, the positivity of the coefficient matrix is automatically satisfied. When the system absorbs energy $\Delta E$ and charge $\Delta Q$, the second law of thermodynamics states that the total entropy change is positive:
\(
\Delta S = \frac{\Delta E - \mu \Delta Q}{T} \geq 0.
\)
Hence, we require that the response of the environment to a weak time-periodic perturbation $H'(t)$ must, on average, produce entropy. 
We define the entropy production
\(
\Sigma(t) \equiv \frac{d}{dt} \frac{\langle H(t) - \mu Q \rangle}{T},
\)
where $H(t)=H_E+H'(t)$. The perturbation is given by
\(
H'(t) = \sum_{i} \left( \lambda_i e^{-i\omega t} B_n(\vb r_i) + \text{H.c.} \right),
\)
where operator $B_n$ satisfies $[Q, B_n] = nB_n$. 
The second law requires that, averaged over one cycle, the entropy production is positive
\begin{align}
    \bar \Sigma = \frac{\omega}{2\pi}\int^{2\pi/\omega}_0\ \mathrm{d}t \,\Sigma(t)\geq 0.
\end{align}
We first calculate the energetic contribution
\begin{align}
\frac{\mathrm{d}}{\mathrm{d}t} \langle H(t) \rangle = \langle \partial_t H'(t) \rangle 
= -i\omega e^{-i\omega t} \sum_{i} \lambda_i \langle B_n(\vb r_i) \rangle + i\omega e^{i\omega t} \sum_{i} \lambda_i^* \langle B_n^\dagger(\vb r_i) \rangle,
\end{align}
where we use the Kubo formula,
\(
\langle B^\dagger_n(\vb r_i) \rangle = \sum_j \lambda_j e^{-i\omega t} G^R[\omega;B^\dagger_n(\vb r_i),B_{n}(\vb r_j)] + \sum_j \lambda^*_j e^{i\omega t} G^R[\omega;B^\dagger_n(\vb r_i),B^\dagger_{n}(\vb r_j)].
\)
The Kubo response of $\langle B_n(\vb r_i)\rangle$ directly follows from  $\langle B_n(\vb r_i) \rangle = \langle B^\dagger_n(\vb r_i) \rangle^*$. We then obtain the time average
\begin{align}
\overline{\frac{\mathrm{d}}{\mathrm{d}t} \langle H(t) \rangle}
&= -i\omega \sum_{ij} \lambda_i \lambda_j^* G^A[\omega;B^\dagger_n(\vb r_j),B_{n}(\vb r_i)] + i\omega \sum_{ij} \lambda_i^* \lambda_j G^R[\omega;B^\dagger_n(\vb r_i),B_{n}(\vb r_j)] \\
&= \omega \sum_{ij} \lambda_i^* \lambda_j \mathcal{A}[\omega;B^\dagger_n(\vb r_i),B_{n}(\vb r_j)].
\end{align}
Here $\mathcal{A}(\omega; A,B) = i[G^R(\omega; A,B) - G^A(\omega; A,B)]$ is the spectral density. Next we calculate the  contribution from the conserved charge $Q$,
\begin{align}
\frac{\mathrm{d}}{\mathrm{d}t} \langle Q \rangle = i\langle [H'(t),Q] \rangle.
\end{align}
After time averaging, we obtain similarly to above
\begin{align}
  \overline{\frac{\mathrm{d}}{\mathrm{d}t} \langle Q \rangle }= n\mu \sum_{i,j}\lambda_i^* \lambda_j \mathcal{A}[\omega;B^\dagger_n(\vb r_i),B_{n}(\vb r_j)].
\end{align}
Thus, the time averaged entropy production is
\(
\bar \Sigma = \frac{\overline{\frac{\mathrm{d}}{\mathrm{d}t} \langle H(t) \rangle} - \mu \overline{\frac{\mathrm{d}}{\mathrm{d}t} \langle Q \rangle }}{T} =  \frac{\omega-n\mu}{T} \sum_{i,j} \left( \lambda_i^* \lambda_j \mathcal{A}[\omega;B^\dagger_n(\vb r_i),B_{n}(\vb r_j)] \right).
\)
The second law of thermodynamics then imposes $\bar \Sigma \geq 0$ which becomes
\(
\sum_{ij}{\lambda}_i^* \Big\{(\omega-n\mu)\mathcal{A}[\omega;B^\dagger_n(\vb r_i),B_{n}(\vb r_j)]\Big\}  {\lambda}_j\geq 0.
\)
Hence, $(\omega-n\mu)\mathcal{A}[\omega;B^\dagger_n(\vb r_i),B_{n}(\vb r_j)]$ is PSD. 

Since the environment is in thermal equilibrium with a nonzero chemical potential $\mu$, the fluctuation-dissipation theorem yields,
\(
G^<[\omega;B^\dagger_n(\vb r_i), B_n(\vb r_j)] = \frac{i\mathcal{A}[\omega;B^\dagger_n(\vb r_i),B_{n}(\vb r_j)]}{1 - e^{\beta(\omega-n\mu)}}, \quad \text{and} \quad
G^>[\omega;B^\dagger_n(\vb r_i), B_n(\vb r_j)] = -\frac{i\mathcal{A}[\omega;B^\dagger_n(\vb r_i),B_{n}(\vb r_j)]}{1 - e^{-\beta(\omega-n\mu)}}, 
\)
where we adopt the standard definitions for lesser and greater Green's functions $G^>(\omega;A,B) = -i\int_{-\infty}^{\infty} \langle A(t)B \rangle e^{i\omega t} dt$ and $G^<(\omega;A,B) = -i\int_{-\infty}^{\infty} \langle BA(t) \rangle e^{i\omega t} dt$. Our notation for environment correlations matches the Green's functions by $C^+_n(\Delta,r_{ij})=iG^>[\omega;B^\dagger_n(\vb r_i),B_{n}(\vb r_j)]$ and $C^-_{n}(\Delta,r_{ij})=iG^<[\omega;B^\dagger_n(\vb r_i),B_{n}(\vb r_j)]$, so that,
\(
C^-_{n}(\Delta,r_{ij}) = -\frac{\mathcal{A}[\omega;B^\dagger_n(\vb r_i),B_{n}(\vb r_j)]}{1 - e^{\beta(\omega-n\mu)}}, \quad \text{and} \quad
C^+_n(\Delta,r_{ij}) = \frac{\mathcal{A}[\omega;B^\dagger_n(\vb r_i),B_{n}(\vb r_j)]}{1 - e^{-\beta(\omega-n\mu)}}.
\)
Since $(\omega-n\mu)\mathcal{A}[\omega;B^\dagger_n(\vb r_i),B_{n}(\vb r_j)]$ is PSD, it follows that $C^\pm_{n}(\Delta,r_{ij})$ are PSD. And because $\abs{\lambda_n}^2\geq0$, the coefficient matrix of the Lindblad equation is guaranteed to be PSD.

\section{Steady state concurrence}

The concurrence of a quantum state $\rho$ can be found by,
\(
\mathcal{C}(\rho)=\text{max}\{0,\lambda_1-\lambda_2-\lambda_3-\lambda_4\},
\)
where $\lambda_i$ is the square root of the eigenvalues of the matrix $\rho(\sigma_y\otimes\sigma_y)\rho^*(\sigma_y\otimes\sigma_y)$ in descending order~\cite{PhysRevLett.80.2245}.
For a two-qubit system, the steady state of the Lindblad equation in Eq. \eqref{eq:Lindblad_diagonal} admits the exact solution with the form
\(\label{eq:rho_solution}
\bar{\rho}_{S}=
\begin{bmatrix}
\rho_{11} &  &  & \\
 & \rho_{22} & \rho_{23} & \\
 & \rho_{32} & \rho_{33} & \\
 & & & \rho_{44}
\end{bmatrix}.
\)
This steady state is unique for the parameter regimes considered.
For such a density matrix, the expression for its concurrence is reduced to
\(
\mathcal{C}(\rho)=2\,\text{max}\{0,\abs{\rho_{23}}-\sqrt{\rho_{11}\rho_{44}}\}.
\)
While this model is analytically solvable, the resulting expressions are cumbersome and not particularly illuminating. For clarity, we therefore present the results in graphical form.

\begin{figure}[h!]
    \centering
    \includegraphics[width=0.95\textwidth]{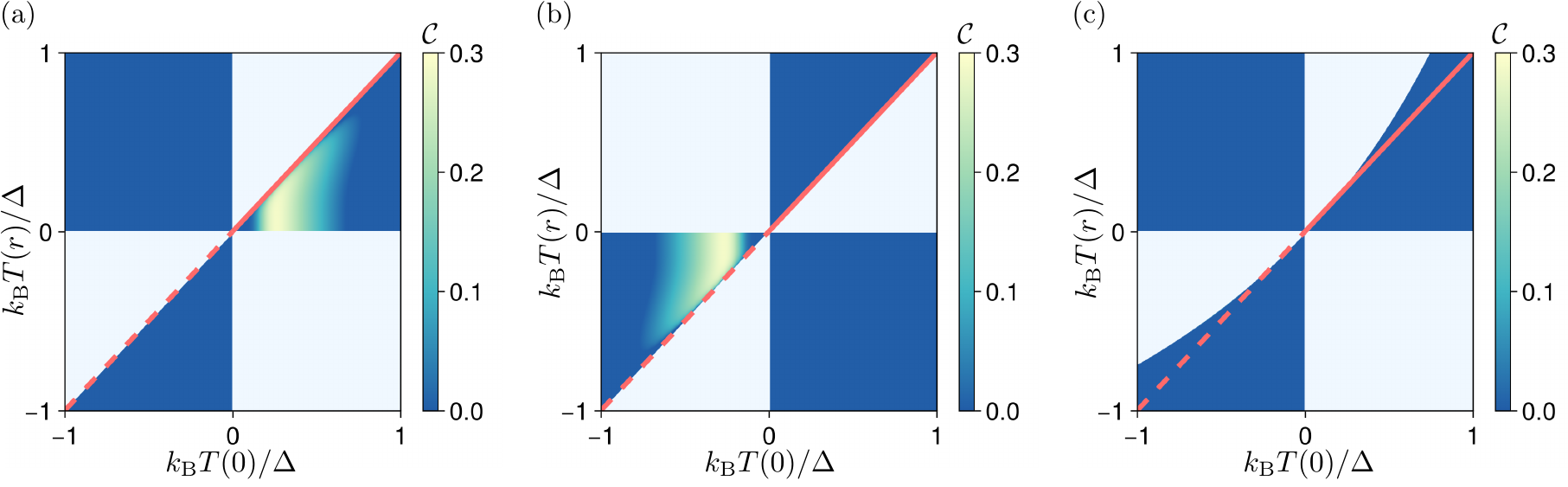}
    \caption{Steady-state concurrence $\mathcal{C}$ of a two-qubit system, as a function of $T(0)$ and $T(r)$, for (a) $\abs{\Gamma_e(r)}/\Gamma_e(0)=0.99$, (b) $\abs{\Gamma_a(r)}/\Gamma_a(0)=0.99$, and (c) $\abs{\Gamma_e(r)}/\Gamma_e(0)=0.7$. The light‐blue regions indicate the unphysical parameters. The red solid (dotted) line marks the positive (negative) temperature equilibrium condition T(0)=T(r).}
    \label{fig:sm_1}
\end{figure}

In Fig.~\ref{fig:sm_1}, we show that entangled steady states arise symmetrically from emission and absorption when $\abs{\Gamma_e(r)}/\Gamma_e(0) \approx 1$ and $\abs{\Gamma_a(r)}/\Gamma_a(0) \approx 1$, occurring only in the quadrant where both local and nonlocal processes are dominated by the same type. In these plots, we fix the relative phase between local and nonlocal rates to $\pi$, i.e., $\Gamma_e(r) = \abs{\Gamma_e(r)}$ and $\Gamma_a(r) = \abs{\Gamma_a(r)}$, as this choice does not qualitatively affect the general conclusion. The constraints on system rates generate multiple `unphysical' regions in the plots, where the corresponding $T(r)$ and $T(0)$ values violate complete positivity of the Lindbladian (i.e., they would require $|\Gamma_e(r)|>\Gamma_e(0)$ or $|\Gamma_a(r)|>\Gamma_a(0)$, in conflict with the PSD conditions derived in the previous section).

\begin{figure}[h!]
    \centering
    \includegraphics[width=0.9\textwidth]{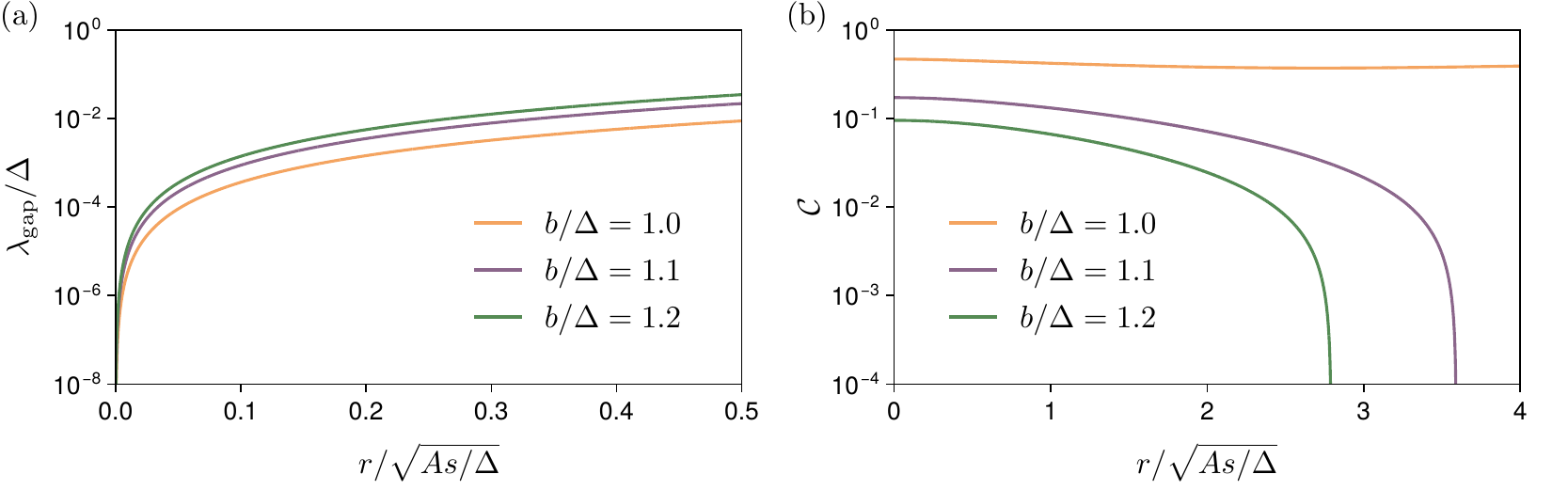}
    \caption{(a) Liouvillian spectral gap $\lambda_\text{gap}/\Delta$ and (b) Concurrence $\mathcal{C}$ as functions of qubit separation $r/\sqrt{As/\Delta}$ for $b/\Delta=1.0$, $1.1$, and $1.2$.}
    \label{fig:sm_2}
\end{figure}

While we have obtained the entangled steady‐state solution, the timescale required to reach it remains to be determined. This timescale is set by the convergence rate of the Lindbladian—often referred to as the Liouvillian spectral gap—which quantifies the asymptotic relaxation speed toward the steady state. The gap is between the steady states and the first eigenvalue with nonzero real part, and therefore determines the slowest relaxation timescale among all transient modes. A vanishing gap corresponds to critical slowing down, while a finite gap guarantees exponential convergence. We numerically evaluate the convergence rates for the spintronics model introduced in the main text, with the results shown in Fig.~\ref{fig:sm_2}(a). The rate exhibits a sharp decline for $r/\sqrt{As/\Delta} \lesssim 0.1$ and follows the general trend that higher concurrence corresponds to slower convergence. However, as shown in Fig.~\ref{fig:sm_2}(b), the exponential decay in the concurrence does not occur until $r/\sqrt{As/\Delta} \gg 1$. Thus, placing the qubit separation in the range $0.1 \lesssim r/\sqrt{As/\Delta} \lesssim 2$ yields near‐maximal concurrence without incurring an impractically long convergence time to reach the steady state.

\section{Detailed Balance in a spatially resolved Lindblad equation}

Detailed balance expresses a microscopic reversibility condition between absorption and emission processes mediated by a thermal reservoir. In a spatially extended system, this reversibility is a property of the reservoir and should hold independently of the separation between any two constituents. 
Here we show detailed balance, for a Lindbladian of general form $\mathcal{L}[\rho] = \sum_{\omega}\sum_{ij}\mathcal{L}^{(\omega)}_{ij}[\rho]$, where
\(
\mathcal{L}^{(\omega)}_{ij}[\rho] =  \gamma_{ij}(\omega) \left( A(\omega,\vb r_i) \rho A^\dagger(\omega,\vb r_j) - \frac{1}{2} \{ A^\dagger(\omega,\vb r_j) A(\omega,\vb r_i), \rho \} \right).
\)
Here, the system operators $A(\omega,\vb r_i)$ are both spatially and spectrally resolved as they act locally on the $i$th site and satisfy $[H_S, A(\omega,\vb r_i)] = \omega A(\omega,\vb r_i)$.
To retrieve the Lindbladian from Eq.~\eqref{eq:Lindblad_superoperator}, we note that there are only two excitation energies $\omega\in\{+\Delta,-\Delta\}$ Then, we can identify $A(-\Delta,\vb r_i)=\sigma^-_i$ and $A(+\Delta,\vb r_i)=\sigma^+_i$ as well as $\gamma_{ij}(-\Delta)=\Gamma_e(r_{ij})$ and $\gamma_{ij}(+\Delta)=\Gamma_a(r_{ij})$.

We now show that the thermal Gibbs state $\rho_T = e^{-\beta H_S}/Z$ satisfies a detailed balance,
\(\label{eq:db}
\mathcal{L}^{(\omega)}_{ij}[\rho_T] + \mathcal{L}^{(-\omega)}_{ji}[\rho_T] = 0 \quad \forall i,j,
\)
when the damping rates satisfy,
\( \label{eq:db2}
\gamma_{ij}(\omega) = e^{-\beta \omega} \gamma_{ji}(-\omega).
\)
For brevity, we denote $A(\omega,r_i) \equiv A_i(\omega)$. 
Using the relations
$[H_S, A_i^\dagger(\omega)A_j(\omega)]=0$ and $e^{\beta H_S} A_i(\omega)e^{-\beta H_S} = e^{\beta \omega} A_i(\omega)$,
we obtain
\(\al{\label{eq:lij}
\mathcal{L}^{(\omega)}_{ij}[\rho_T]
&= \gamma_{ij}(\omega) \rho_T \left[
e^{\beta \omega}  A_i(\omega) A_j^\dagger(\omega)
-  A_j^\dagger(\omega) A_i(\omega)
\right].
}\)
Similarly, we evaluate $\mathcal{L}^{(-\omega)}_{ji}[\rho_T]$,
\(\al{\label{eq:lji}
\mathcal{L}^{(-\omega)}_{ji}[\rho_T]
&= \gamma_{ji}(-\omega)e^{-\beta \omega} \rho_T \left[
  A_j(-\omega) A_i^\dagger(-\omega)
-  e^{\beta \omega} A_i^\dagger(-\omega) A_j(-\omega)
\right].
}\)
After identifying $A_i(-\omega) = A_i^\dagger(\omega)$~\cite{Breuer} and adding both contributions, we see that,
\(\al{
\mathcal{L}^{(\omega)}_{ij}[\rho_T]+\mathcal{L}^{(-\omega)}_{ji}[\rho_T]
&= \left[\gamma_{ij}(\omega)-\gamma_{ji}(-\omega)e^{-\beta \omega}\right] \rho_T \left[
e^{\beta \omega}  A_i(\omega) A_j^\dagger(\omega)
-  A_j^\dagger(\omega) A_i(\omega)
\right].
}\)
Hence, if the relation in Eq.~\eqref{eq:db2} is fulfilled, detailed balance according to Eq.~\eqref{eq:db} is satisfied for all $i,j$ and $\omega$, thereby establishing the exact cancellation of emission and absorption contributions. 

\section{Breaking conservation law with exchange coupling}
Consider a system of $N$ independent qubits of energy $\Delta$ subject to an external field, whose interaction with a spintronic environment is through an exchange coupling  $H_I=-J \sum_i^N \bm{\sigma}_i\cdot\vb{S}_i$, where $\bm{\sigma}_i$ is the Pauli vector of qubit $i$, and $\vb{S}_i$ is the environment spin operator on site $i$. We define the spin of the environment to be in $+z$ direction and the system spin is misaligned by an angle $\theta$ clockwise to the $z$ axis in the $yz$ plane, such that
\(
\sigma^{x}_i = \sigma^{x'}_i, \quad
\sigma^{y}_i = \sigma^{y'}_i \cos{\theta} - \sigma^{z'}_i \sin{\theta}, \quad
\sigma^{z}_i = \sigma^{y'}_i \sin{\theta} + \sigma^{z'}_i \cos{\theta},
\)
where $x',y',z'$ describes the system frame. The interaction Hamiltonian can be expressed by
\(\al{
H_I 
&=-J\sum_i^N \ \sigma^x_i S^x_i + \sigma^y_i S^y_i + \sigma^z_i S^z_i \\
&=-J\sum_i^N \ \sigma^{x'}_i S^x_i + \cos{\theta}\sigma^{y'}_i S^y_i + \sin{\theta}\sigma^{y'}_i S^z_i - \sin{\theta}\sigma^{z'}_i S^y_i + \cos{\theta}\sigma^{z'}_i S^z_i.
}\)
We discard the dephasing ($\sigma^{z'}_i$) terms, and drop the term involving $\sigma^{y'}_i S^z_i$, which results in a static field term that shifts $H_S$. We also omit higher order terms in the large spin approximation. The resultant Hamiltonian is
\(\al{
H_I 
&=-J\sum_i^N\ \sigma^{x'}_i S^x_i + \cos{\theta}\sigma^{y'}_i S^y_i \\
&=-J\sum_i^N\ (1-\cos{\theta})\big(\sigma^+_iS^+_i+\sigma^-_iS^-_i\big)+(1+\cos{\theta})\big(\sigma^+_iS^-_i+\sigma^-_iS^+_i\big),
}\)
where $\sigma^\pm_i = (\sigma^{x'}_i\pm i\sigma^{y'}_i)/2$ and $S^\pm_i = \big(S^x_i\pm iS^y_i\big)/2$. This Hamiltonian resembles $H_I$ in the main article: in the rotated qubit basis one can identify $\lambda_{1} = -J(1+\cos\theta)$ and $\lambda_{-1} = -J(1-\cos\theta)$,
so that the spin-conserving and spin-nonconserving channels correspond directly to the terms in the main text.

\section{Correlation functions of the inverted magnet}
The two-dimensional magnetic environment is described by Hamiltonian,
\( 
H_E=\int \mathrm{d}^2r\,\frac{A}{2}[\nabla \vb{s}]^2 - bs_z,
\)
where $\vb s(\vb r)$ is the spin density operator, $A$ is the spin stiffness of the ferromagnet, and $b=\gamma_e B>0$ describes an external field $B<0$ pointing in the $-\hat{z}$ direction, with gyromagnetic ratio $\gamma_e<0$. 
The environment is prepared in an \textit{inverted} configuration, with spins polarized opposite to the applied field and described by the spin grand-canonical ensemble
\(
\bar\rho_E = e^{-(H_E - \mu S_z)/k_\text{B} T}/Z, \quad
S_z = \int \mathrm{d}^2r\, s_z(\vb r),
\)
with spin accumulation $\mu<-b$ and $T\rightarrow 0$, so that the reference state is the fully polarized state $\ket{\downarrow\downarrow\dots}$. 
While both $\mu$ and $b$ favor a certain spin direction, we would like to refute a frequently encountered but incorrect claim that they are interchangeable. The applied magnetic field $b$ modifies the dynamics of the spin system as it sets the Larmor precession and the magnon dispersion. By contrast, the spin accumulation $\mu$ characterizes the statistical ensemble of the environment and does not modify the precession frequency. Moving on, we linearize the Hamiltonian by applying the Holstein-Primakoff transformation with respect to this inverted state: 
\(
s_+(\vb r) = s_-^\dagger(\vb r) \approx \sqrt{2s}a^\dagger(\vb r),\quad
s_z(\vb r)=-s+a^\dagger(\vb r)a(\vb r), 
\)
where $a^\dagger, a$ are magnon operators satisfying the bosonic commutation relation and $s$ is the saturated spin density. In this convention, the inverted state $\ket{\downarrow\downarrow\dots}$ is the bosonic vacuum for $a(\vb r)$. The exchange term can be written as
\(\al{
[\nabla \vb{s}]^2 
\approx \frac{1}{2} \partial_\alpha s_+ \partial_\alpha s_- + \frac{1}{2} \partial_\alpha s_- \partial_\alpha s_+ 
= 2s \, (\nabla a^\dagger) \cdot (\nabla a),
}
\)
where we have dropped higher-order terms in $a$ and $a^\dagger$. The Zeeman term becomes
\(
-bs_z =-b(-s+a^\dagger a) = bS - b a^\dagger a.
\)
Finally, we find the transformed Hamiltonian,
\(
H_E = \int \mathrm{d}^2 r \, [As(\nabla a^\dagger(\vb r))(\nabla a(\vb r)) - b a^\dagger(\vb r) a(\vb r)] + \text{const.
}
\)
In momentum space, the Hamiltonian reads
\( 
H_E=\int {\mathrm{d}^2k}\,(As\vb{k}^2-b)a^\dagger_{\vb{k}} a_{\vb{k}},
\)
where we have performed the Fourier transformation $a(\vb r)=\int \mathrm{d}^2k\,e^{-i\vb{k}\cdot\vb{r}}a_{\vb k}/2\pi$. 
The dispersion is given by $\omega(\vb k) = As\vb k^2 - b$, with a minimum at $\omega_\text{min} = -b$. Excitations with $\omega_{\vb{k}}<0$ lower the energy of the inverted reference state and are also referred to as antimagnons.
The explicit expression for the correlation functions of the environment can then be carried out, e.g., for $C^+_1(\omega,r)$,
\(\al{
C^+_{1}(\omega,r_{ij}) &= \int_{-\infty}^\infty \mathrm{d}t \ e^{i\omega t}\langle s_-(\vb{r}_i,t)s_+(\vb{r}_j) \rangle \\
& = 4\pi s \int_0^\infty \frac{k \, \mathrm{d}k}{(2\pi)^2} \int_0^{2\pi} \mathrm{d}\theta \, e^{i k r_{ij} \cos\theta} [1 + n_B(\omega)] \delta(\omega_k - \omega)\\
& = 4\pi s\ [n(\omega)+1]g(\omega)J_0[r_{ij}\sqrt{(b+\omega)/As}],
}\)
where $g(\omega) = \frac{1}{4\pi As}\,\Theta(\omega+b)$ is the density of states, with $\Theta(\dots)$ denoting the Heaviside step function, and $n(\omega) = 1/(\exp [(\omega-\mu)/k_\text{B}T]-1)$ denotes the Bose-Einstein distribution. Similarly, we can write down:
\(
C^+_{-1}(\omega,r_{ij}) = \int_{-\infty}^\infty \mathrm{d}t \ e^{i\omega t}\langle s_+(\vb{r}_i,t)s_-(\vb{r}_j) \rangle = 4\pi s\ [n(-\omega)]g(-\omega)J_0[r_{ij}\sqrt{(b-\omega)/As}].
\)
The occupation $n(\omega)\to0$ when $T\to0$, such that the correlation $C^+_{-1}(\Delta,r)=0$ and does not contribute to $\Gamma_e(r)$. Similarly, $C^-_1(\Delta,r) =0$ and only $C^-_{-1}(\Delta,r) = 4\pi s\ g(-\Delta)J_0[r\sqrt{(b-\Delta)/As}]$ contributes to the absorption process.

\bibliography{citations_sm}